\documentclass[manuscript]{aastex631}
\usepackage{amsmath}

\begin{document}

\title{Reactive molecular dynamics simulations of micrometeoroid bombardment for space weathering of asteroid (162173) Ryugu}

\correspondingauthor{Daigo Shoji}
\email{shohji.daigo@jaxa.jp}

\author[0000-0001-6423-0698]{DAIGO SHOJI}
\affiliation{Institute of Space and Astronautical Science, Japan Aerospace Exploration Agency, Chuo-ku, Sagamihara 252-5210, Japan}




\begin{abstract}
Remote sensing observations by Hayabusa2 and laboratory measurements have revealed that the phyllosilicates on asteroid (162173) Ryugu are dehydrated/dehydroxylated due to space weathering. Reactive molecular dynamics simulations were performed to evaluate the magnitude of the dehydroxylation of Mg-rich serpentine by micrometeoroid impacts. When micrometeoroids were not coupled with interplanetary magnetic fields, serpentine could be dehydroxylated by micrometeoroids as small as 2 nm in size. In particular, $\sim$200 O-H bonds dissociated when the meteoroids were derived from cometary activity (the impact velocity was $\sim$20 km s$^{-1}$). When nano-sized dust particles were accelerated to $\sim$300 km s$^{-1}$ by the magnetic fields of solar wind plasma, the number of dissociated O-H bonds increased by one order of magnitude. Consequently even 1 nm-sized dust particles can contribute to the space weathering of Ryugu. In all cases, Si-OH, H$_2$O, and free OH were generated from the hydroxyls initially connected to Mg, which could partially offset dehydration. Despite the limitations of our computational resources, which restricted the simulation time scale to 1 ps, reactive molecular dynamic simulations demonstrated that micrometeoroid bombardment could influence the space weathering of asteroids. 

\end{abstract}



\section{Introduction} \label{sec:intro}
From 2018 to 2019, the Hayabusa2 spacecraft, which was launched by the Japan Aerospace Exploration Agency (JAXA), approached the near-earth asteroid (162173) Ryugu and conducted in-situ observations \citep[e.g.,][]{sugita2019, kitazato2019, watanabe2019}. During these observations, the near-infrared spectrometer (NIRS3) detected a $\sim$2.7 $\mu$m absorption in the reflectance spectrum, which corresponds to the O-H stretching vibration mode \citep{kitazato2019}. This observation is evidence that the surface of Ryugu has Mg-rich phyllosilicates, such as serpentine and saponite \citep{sugita2019, kitazato2019}. In April 2019, Hayabusa2 created an artificial impact crater, and the geology at $\sim$1 m subsurface was also observed. Compared to the surface, the $\sim$2.7 $\mu$m absorption feature was slightly stronger in the crater \citep{kitazato2021}. This difference in absorption features implies that the surface of Ryugu is affected by space weathering, which results in the dissociation of hydroxyls \citep{kitazato2021}. 

After the in-situ observations, Hayabusa2 returned samples from Ryugu to Earth. The $\sim$2.7 $\mu$m absorption band was also identified in the returned samples through laboratory measurements \citep{yada2022, pilorget2022, pivert-jolivet2023}. However, the depth of the absorption band was twice as large as that observed through remote sensing \citep{yada2022, pilorget2022, pivert-jolivet2023}. To resolve this discrepancy, \citet{matsuoka2023} measured the spectrum of the Murchison meteorite, a compositional analog for Ryugu, with different porosities and grain sizes. While particle size affects the $\sim$2.7 $\mu$m absorption feature, they found that the weak absorption band identified by remote sensing observations seemed to be caused by space weathering, especially space weathering by micrometeoroid bombardment. Moreover, laboratory experiments that simulated micrometeoroid impacts on Murchison using laser pulses observed changes in the absorption band at $\sim$3.0 $\mu$m \citep{matsuoka2015, matsuoka2020}.

In addition to micrometeoroid bombardment, charged particles, such as protons from the solar wind, also play an important role in space weathering, as implied from sample analyses \citep{hiroi2023}. However, the saturation timescale of solar wind protons is only $\sim$$10^2 - 10^3$ years, which is much shorter than the resurfacing timescale of Ryugu \citep{nakauchi2021, matsuoka2023}. Moreover, laboratory experiments in which protons were irradiated to phyllosilicate indicated that the protons in the solar wind tend to increase the number of hydroxyls by breaking the Si-O-Si bonds within phyllosilicate and binding with the oxygen atoms \citep{nakauchi2021}. Thus micrometeoroid bombardment can be one of the most important processes causing the dehydration/dehydroxylation of Ryugu's surface. 

Although laboratory experiments are crucial for understanding space weathering on asteroids, it is difficult to observe detailed interactions of atoms because chemical reactions occur within sub-nanoseconds. To evaluate the process of space weathering at an atomic scale, molecular dynamics (MD) simulations can be useful. Previous MD simulations using reactive force fields (ReaxFF), which can simulate bond changes, have been performed for lunar and icy bodies' surface compositions \citep{huang2021, huang2022, anders2013, anders2017, anders2019, grayson2022}. However, reactive MD simulation for space weathering on an asteroid's surface has not been conducted thus far.

In this study, MD simulations were performed for micrometeoroid bombardment on antigorite (Mg-bearing phyllosilicate) to evaluate the magnitude of OH reduction. Using the Large-scale Atomic/Molecular Massively Parallel Simulator (LAMMPS) \citep{thompson2022} and the ReaxFF potential, the change in the amount of OH and H$_2$O was calculated. Due to our computational resources, the simulations in this work have limitations in spatial and time scales. However, despite these limitations, the MD simulations with different parameters can evaluate the magnitude of OH reduction by micrometeoroids.

\section{Method} \label{sec:style}
\subsection{Reactive force field}
In MD simulation, each atom is treated as a particle influenced by a force field. Under an assumed force field, atoms can move and interact with each other. Among various potential models, this study employs the reactive force field (ReaxFF) potential, which was developed by \citet{van2001}. The most distinctive feature of ReaxFF is that it can simulate chemical reactions (bond changes) between atoms. 

The ReaxFF potential $E_{Reax}$ between atoms is given as the sum of several terms as follows:

\begin{equation}
E_{Reax} = E_{bond} + E_{lp} + E_{over} + E_{under} + 
 E_{val} + E_{pen} + E_{coa} + E_{tors} + E_{conj} + E_{hbond} + E_{vdWaals} + E_{Coulomb}.
\end{equation}
Each term represents the bonding energy $E_{bond}$, lone-pair energy $E_{lp}$, overcoordination energy $E_{over}$, undercoordination energy $E_{under}$, valence-angle energy $E_{val}$, penalty energy $E_{pen}$, three-body conjugation energy $E_{coa}$, torsion-angle energy $E_{tors}$, 4-body conjugation energy $E_{conj}$, and hydrogen bonding energy $E_{hbond}$. $E_{vdWaals}$ and $E_{Coulomb}$ are the non-covalent interactions of van der Waals energy and Coulomb energy \citep{huang2021, huang2022}. 
LAMMPS can implement ReaxFF using a potential file. In this study, antigorite (serpentine), which consists of Si, Mg, O, and H, was considered as the target, and the impactors (micrometeoroids) were assumed to be composed of silica for simplicity. Recently, \citet{yeon2023} evaluated the ReaxFF potentials of magnesium
aluminosilicate glass with hydroxyls. Here, their potential data were used in the simulations. 

\subsection{Simulation domain}
\begin{figure}[ht!]
\plotone{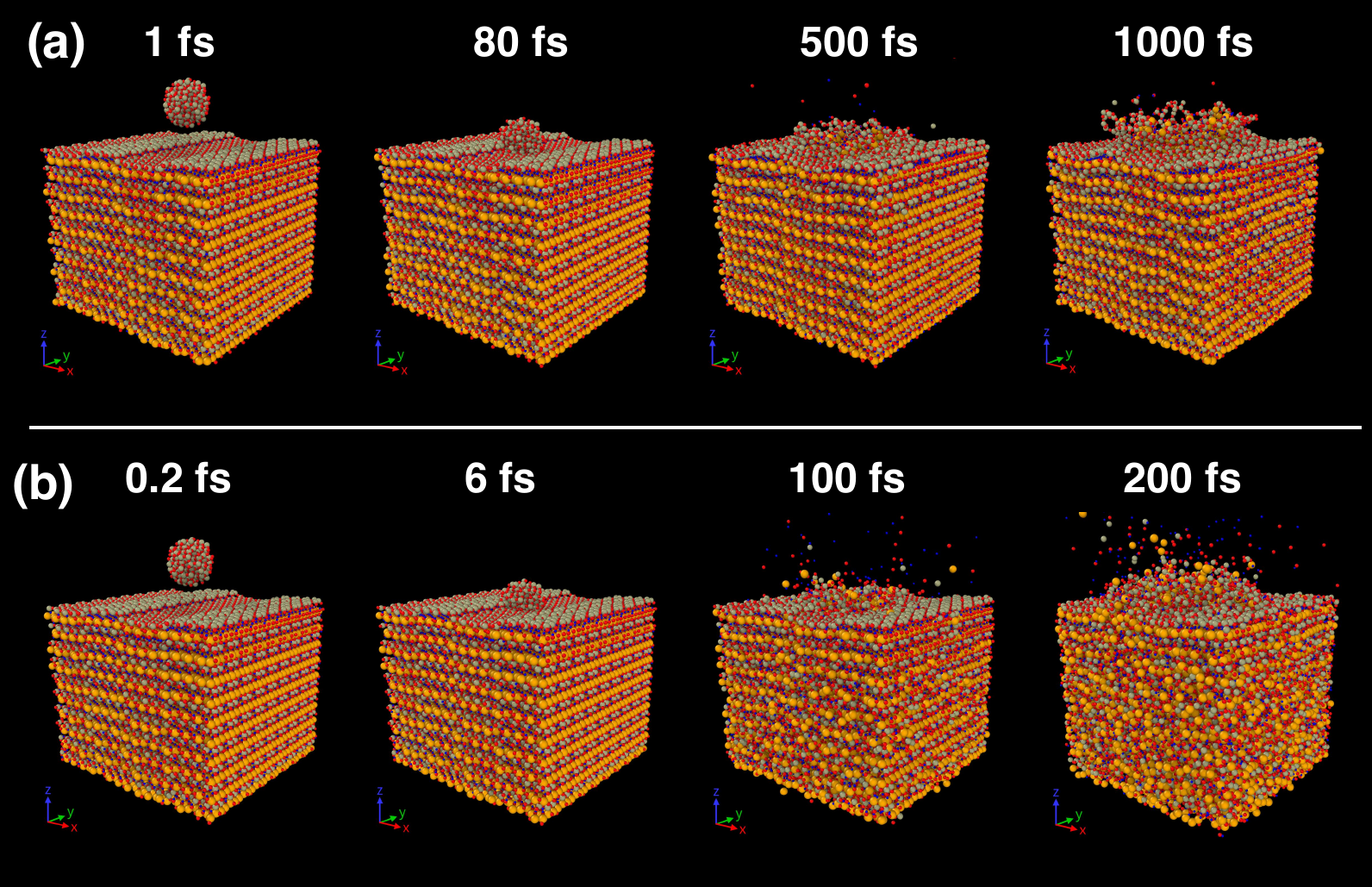}
\caption{Snapshots of the simulations at $D$ = 2 nm, and $v$ = 20 km s$^{-1}$ (a) and $v$ = 300 km s$^{-1}$ (b). The colors of each particle represent Mg (yellow), Si (gray), O (red), and H (blue), respectively.}
\label{snap}
\end{figure}

An antigorite block was used as the target domain (Fig. \ref{snap}) as spectral analyses in the mid-infrared region of the Ryugu samples revealed that antigorite is one of the main phyllosilicate phases present in the returned samples \citep{yesiltas2024}. The block's dimensions were approximately 8.7 nm $\times$ 8.2 nm $\times$ 7.3 nm in the XYZ direction. Although LAMMPS can handle larger blocks, this size was chosen based on computational resource constraints. The basal structure of the block was obtained from \citet{capitani2004}, whose data are available in the Crystallography Open Database (COD). Periodic boundary conditions were applied in the X- and Y-directions. The structure of the SiO$_2$ impactor was obtained from \citet{glinnemann1992}. The diameter of the impactor $D$ was set to 1 nm and 2 nm considering the $\sim$8 nm size of the target. The impactors were initially positioned at the center of the target's X-Y plane (Fig. \ref{snap}) above 6 \AA~ from the surface  (i.e., the impactor's center was above 0.5$D$ + 6 \AA). In LAMMPS, the default cutoff distance for the bond interactions of ReaxFF is 5 \AA. Thus, at the initial condition, no interactions occurred between the impactor and the target.

\subsection{Impact velocity}
To determine the impact velocity of micrometeoroids on asteroids, \citet{altobelli2019} conducted numerical simulations using the ESA Interplanetary Meteoroid Engineering Model (IMEM) \citep{dikarev2005}. In the case of the near-earth asteroid Eros (its semi-major axis is 1.45 AU), micro dust particles caused by asteroid collisions have approximately 10 km s$^{-1}$ impact velocity while velocities due to cometary activity range between $\sim$10 km s$^{-1}$ and $\sim$30 km s$^{-1}$, depending on the true anomaly \citep{altobelli2019}. In this study, for simplicity, simulations were performed with impact velocities $v$ of 10 km s$^{-1}$ and 20 km s$^{-1}$, which correspond to the average velocities of dust particles from asteroid collisions and cometary activity, respectively. 

One caveat is that these impact velocities assume meteoroids are not coupled with the interplanetary magnetic field of solar wind plasma. Due to the charge of micrometeoroid and the magnetic field of solar wind, a Lorentz force is induced, which can easily accelerate micrometeoroids with small radii due to their lower mass \citep{juhasz2013}. Because \citet{altobelli2019} considered meteoroids of micrometers in size, they did not account for the effect of the interplanetary magnetic field. For micrometeoroids larger than 0.2 $\mu$m, the interplanetary magnetic field does not significantly affect the impact speed \citep{altobelli2019}. In contrast, when the dust particle size is a few nanometers, coupling with the interplanetary magnetic field of solar wind plasma accelerates the dust particle to $\sim$300 km s$^{-1}$ of velocity \citep[e.g.,][]{juhasz2013, poppe2020}. Due to computation cost, the diameters of the impactors in this study were assumed to be 1 nm and 2 nm. Thus, in addition to the velocities of 10 km s$^{-1}$ and 20 km s$^{-1}$, which can be regarded as low-speed impacts without coupling with the magnetic field, high-speed simulations with $v$ = 300 km s$^{-1}$ were also performed. At this velocity, the kinetic energy of each atom in the impactor exceeds 1 keV. Typically, the ReaxFF potential is evaluated for interactions with kinetic energies below a few tens of electron volts. In several studies on high-energy simulations, in addition to the ReaxFF potential, the Ziegler-Biersack-Littmark (ZBL) potential \citep{ziegler1985} is also considered to simulate repulsive interactions \citep{huang2022, anders2013, anders2017, anders2019}. The ZBL potential between the $i$th and $j$th atoms $E^{ZBL}_{ij}$ can be represented as follows

\begin{equation}
E^{ZBL}_{ij} = \frac{1}{4\pi\epsilon_0}\frac{Z_i Z_j e^2}{r{ij}}\phi(r_{ij}/a),
\end{equation}
where $e$, $\epsilon$, and $Z$ are the electron charge, the electrical permittivity of vacuum, and the nuclear charge of the atom, respectively \citep{ziegler1985, huang2022}. The parameters $\phi(x)$ and $a$ are given as follows,

\begin{equation}
\phi(x)=0.18175e^{-3.19980x}+0.50986e^{-0.94229x}+0.28022e^{-0.40290x}+0.02817e^{-0.20162x},
\end{equation}

\begin{equation}
a = \frac{0.46850}{Z_i^{0.23}+Z_j^{0.23}}.
\end{equation}
This study also considered the ZBL potential when $v$ = 300 km s$^{-1}$. Because our simulation region is limited, high-speed simulations are preliminary; therefore, simulations using larger regions will be performed in the future. However, we evaluated the change in the number of hydroxyls due to the high-speed impact by comparing it with the low-speed simulations. 

\subsection{Simulation setup}
The initial temperature of the target was assumed to be 330 K. Although the surface temperature of Ryugu varies with location and time, the globally averaged maximum temperature is 339$\pm$12 K, and the surface temperature changes diurnally between $\sim$310 and $\sim$340 K \citep{shimaki2020}. Thus, a temperature of 330 K is consistent with the average diurnal surface temperature of Ryugu. When the surface is not exposed to sunlight, the surface temperature decreases to $\sim$200 K \citep{grott2019}. A simulation was conducted at 200 K; however, the number of hydroxyls did not change significantly, as the energy required to induce chemical reactions was given from the kinetic energy of the impactors. 

In each simulation, prior to impact, a canonical ensemble (NVT) simulation, in which the number of particles (N), volume (V), and temperature (T) are conserved, was conducted for 2,000 steps with a time step of 0.1 fs, which made the simulation domain 330 K. Then, the impactor was shot using a microcanonical ensemble (NVE) simulation, where the number of particles (N), volume (V), and energy (E) are conserved. When $v$ = 10 km s$^{-1}$ and 20 km s$^{-1}$, one time step was set to 0.1 fs following \citet{huang2021}. In the high-speed simulations, one time step was reduced to 0.02 fs. A simulation with a smaller time step of 0.01 fs was also performed, and the number of hydroxyls did not change significantly. Each simulation was run for up to 10,000 steps. Thus, interactions were simulated up to 1 ps (1,000 fs) at $v$ = 10 km s$^{-1}$ and 20 km s$^{-1}$, and 0.2 ps (200 fs) at $v$ = 300 km s$^{-1}$. 
In the NVE simulation, the total energy is conserved. Therefore, if the simulations are extended over a long period, the effect of thermal diffusion (the diffusion of atomic motion caused by the impact) could reach the periodic boundaries and re-enter the simulation region from the opposite boundary, which prevents a reasonable temperature decrease. Given the thermal diffusivity of serpentine $\kappa$ at $\sim$$10^{-6}$ m$^2$ s$^{-1}$ \citep{osako2010} and a length scale $l$ of $\sim$1 nm, the characteristic time for thermal diffusion ($l^2$/$\kappa$) is approximately 1 ps. In this study, the NVE simulations were performed before the thermal diffusion effect reached the boundaries. For longer simulations that account for thermal diffusion, a larger volume of simulation region is necessary, which is planned for future work. 

For the calculation of atomic charges, the charge equilibration (QEq) method was applied (see \citet{aktulga2012} for details). However, when $v$ = 300 km s$^{-1}$, many atoms are ejected from the surface and the QEq method does not work correctly. Thus, QEq was turned off in the NVE simulation. In the initial NVT calculations, the QEq method was used even at $v$ = 300 km s$^{-1}$ to achieve charge equilibrium. 

\subsection{Calculation of OH and H$_2$O number}
The number of hydroxyls must be calculated to evaluate the effect of space weathering by each impact. ReaxFF uses bond order as a factor of bond strength between atoms. \citet{yeon2023} evaluated that a bond order greater than 0.65 between O and H atoms can regard O-H as a connected hydroxyl. Thus, to determine the amount of connected OH, the number of O-H pairs with a bond order more than 0.65 was calculated. In addition, the numbers of hydroxyls connected to Mg and Si atoms (i.e., Mg-O-H and Si-O-H) were also calculated following the critical bond orders of 0.31 for Mg-O and 0.8 for Si-O, respectively \citep{yeon2023}. 

In addition to the bond order, LAMMPS can output independent molecules at each step. Thus, the numbers of H$_2$O and free OH were also evaluated. The critical bond order, except for the pairs shown above, was set to 0.3, which is the default value of LAMMPS.

Even though there is no impact, the O-H bond number changes slightly due to the thermal oscillations of atoms (bond order sometimes lowers the critical value). In this work, MD simulations without impact were also performed to evaluate the effect of the impact itself, and the numbers of OH and H$_2$O were calculated as background amounts. Then, the differences between the simulations with and without impact at the same step were regarded as the change caused by one impact event. 

\section{Result and Discussion} \label{sec:floats}
\subsection{Low-speed impact}
Fig. \ref{snap} (a) shows snapshots of an impact event at $v$ = 20 km s$^{-1}$. Due to the impact, a crater was generated, and the surface transitioned to an amorphous state. In the cases of $v$ = 20 km s$^{-1}$, the impactors penetrated the target to depths of $\sim$4.4 nm and $\sim$2.2 nm at $D$ = 2 nm and 1 nm, respectively. The ratio of these penetration depths ($\sim$2) is roughly consistent with the scaling law of penetration depth $p$, given as $p \propto D^{1.056}$ \citep{berthoud1997}. The kinetic energy of the impactor was transferred to the atoms in the target, inducing cascade reactions. Serpentine becomes unstable above 600 $^{\circ}$C \citep{matsuoka2023}. In every simulation, the temperature around the impact point exceeded 1,000 K due to the kinetic energy received from the impactor. 

\begin{figure}[ht!]
\plotone{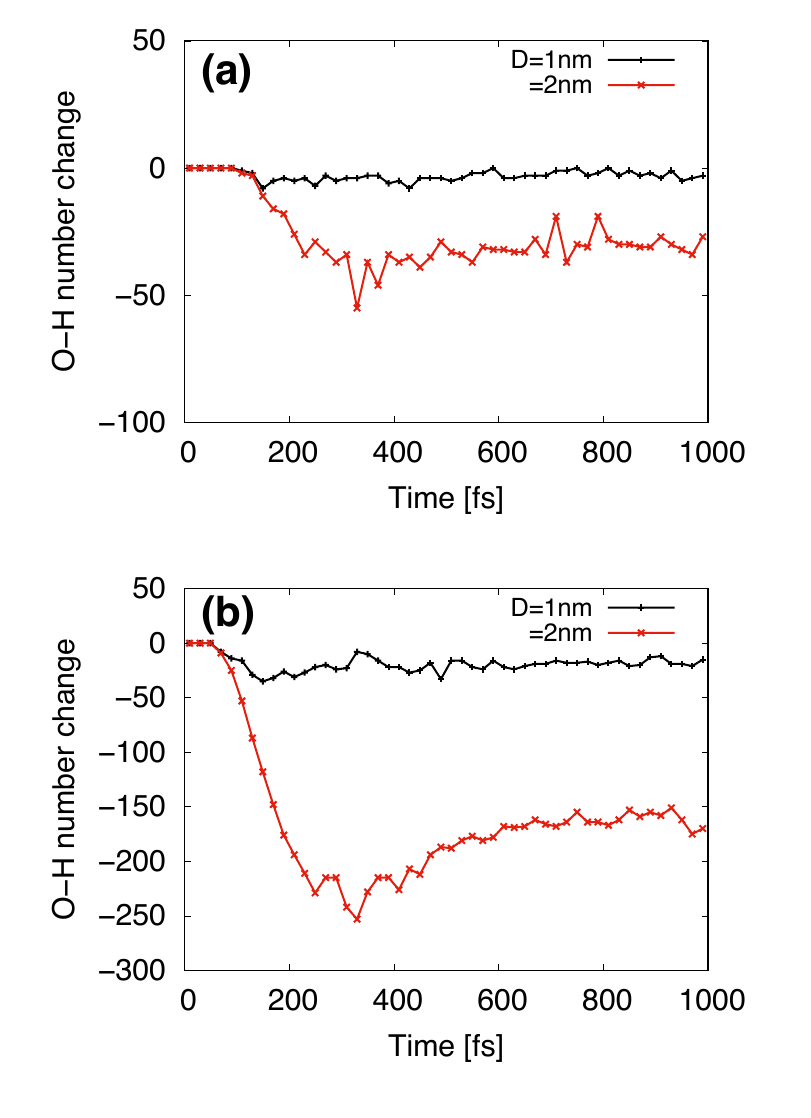}
\caption{Changes in the total O-H bonds at $v$ = 10 km s$^{-1}$ (a) and $v$ = 20 km s$^{-1}$.} 
\label{OH_slow}
\end{figure}

The changes in the total number of O-H bonds (O-H bonds in every molecule and free OH) at $v$ = 10 km s$^{-1}$ and 20 km s$^{-1}$ are shown in Fig. \ref{OH_slow}. In all cases, the number of O-H bonds decreased, which is consistent with the laser experiments that simulated micrometeoroid bombardment \citep{matsuoka2015, matsuoka2020}. As the size and velocity of the impactor increased, more O-H bonds were broken because the large kinetic energy of the impactor dissipated in the target. However, when the diameter was 1 nm, even at $v$ = 20 km s$^{-1}$, the change in the O-H bond number was small. Thus, when a micrometeoroid is not coupled with the interplanetary magnetic fields, a particle size greater than approximately 2 nm is required to significantly reduce the number of OH. Although our simulations assumed nano-sized impactors, if dust particles with a few $\mu$m strike the surface of Ryugu, the effect of space weathering would increase drastically because the kinetic energy depends on the mass of the impactor. 

At $v$ = 20 km s$^{-1}$ and $D$ = 2 nm, the number of O-H bonds slightly increased after 400 fs (Fig. \ref{OH_slow} b). As shown below, this increase can be attributed to the generation of Si-OH, H$_2$O, and free OH from O and H initially connected as Mg-OH, which resulted in an overall increase in the total number of O-H bonds. At $D$ = 1 nm or $v$ = 10 km s$^{-1}$, although these molecules were also generated, the amount was insufficient to cause a clear increase in O-H bonds.

\begin{figure}[ht!]
\plotone{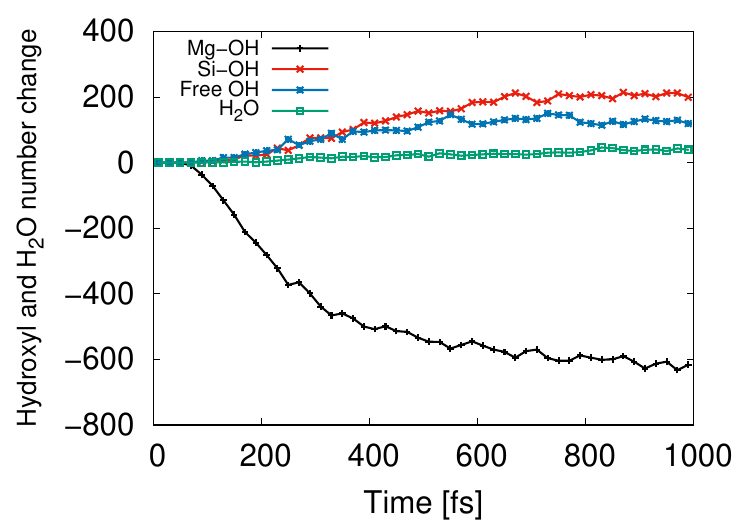}
\caption{Changes in the hydroxyls connected to Mg and Si atoms, free OH, and H$_2$O molecules at $D$ = 2 nm and $v$ = 20 km s$^{-1}$.} 
\label{OH_H2O_slow}
\end{figure}

Fig. \ref{OH_H2O_slow} shows the changes in the number of each hydroxyl type and H$_2$O molecules at $D$ = 2 nm and $v$ = 20 km s$^{-1}$. Because antigorite has hydroxyls only connected to Mg atoms, Mg-O-H bonds were broken at the beginning of the impact event (i.e., one of three cases occurred: Mg-O + H, Mg + O-H, and Mg + O + H). Although the number of Mg-OH further decreased as the impact progressed, Si-OH and free OH began to increase. The number of H$_2$O also increased with time. The bonding energy of six-coordinated Mg-O is 155 kJ mol$^{-1}$, which is much lower than the bonding energies of four-coordinated Si-O (443 kJ mol$^{-1}$) and of O-H (425.8 kJ mol$^{-1}$) \citep{ke2020, blanksby2003}. Consequently, Mg-OH can easily dissociate under impact, leading to the production of free OH along with O and H atoms. These dissociated atoms then recombine, generating Si-OH and H$_2$O. The increases in Si-OH and H$_2$O can offset the decrease in O-H bond.

\subsection{High-speed impact}
Fig. \ref{snap} (b) shows snapshots of an impact event at $D$ = 2 nm and $v$ = 300 km s$^{-1}$, which corresponds to the scenario in which nano-scale dust particles are accelerated by the interplanetary magnetic field of solar wind plasma \citep[e.g.,][]{juhasz2013, poppe2020}. Compared with the low-speed impact events, a significantly greater number of atoms were ejected. Due to the limited simulation domain, the entire region became disordered by 200 fs as a result of cascade interactions. To simulate a more reasonable environment, a larger volume of the simulation region is required. However, we estimated the trend in OH changes based on these simulation results.

\begin{figure}[ht!]
\plotone{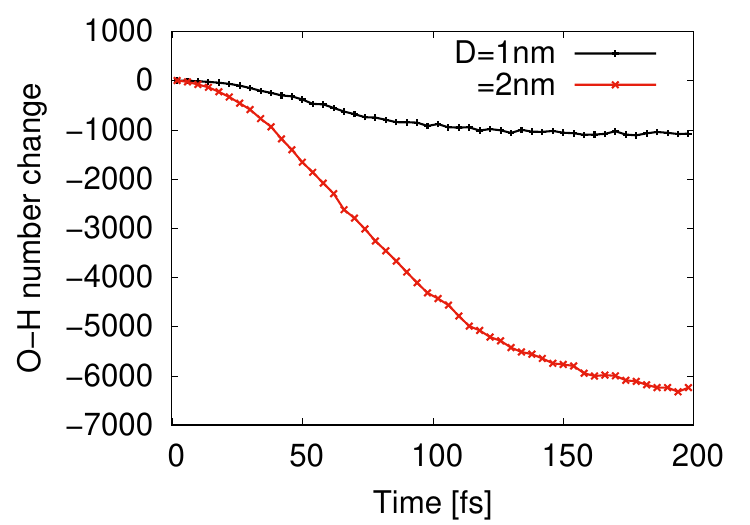}
\caption{Changes in the total O-H bonds at $v$ = 300 km s$^{-1}$.} 
\label{OH_high}
\end{figure}

\begin{figure}[ht!]
\plotone{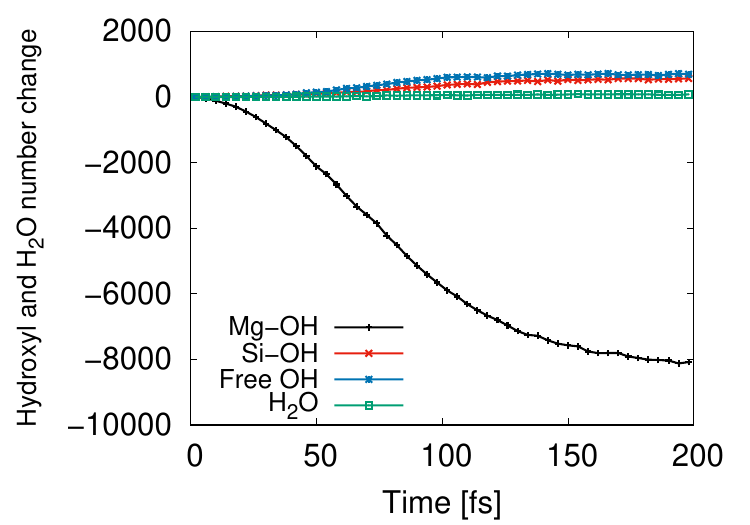}
\caption{Changes in the hydroxyls connected to Mg and Si atoms, free OH, and H$_2$O molecules at $D$ = 2 nm and $v$ = 300 km s$^{-1}$.} 
\label{OH_H2O_high}
\end{figure}

Fig. \ref{OH_high} shows the change in the total number of O-H bonds. Compared with the low-speed simulations, a substantial reduction in O-H bonds was observed. Even at $D$ = 1 nm, more than 1,000 O-H bonds were broken. If a deeper simulation domain and longer steps were considered, even more O-H bonds would be dissociated. The changes in each hydroxyl type and H$_2$O molecules are shown in Fig. \ref{OH_H2O_high}. Si-OH and free OH tended to increase, similar to what was observed in the low-speed simulations. The number of H$_2$O molecules also increased. However, while the magnitude of Mg-OH reduction increased by an order of magnitude at $v$ = 300 km s$^{-1}$ and $D$ = 2 nm compared with the case at $v$ = 20 km s$^{-1}$ and $D$ = 2 nm, the numbers of Si-OH and H$_2$O increased only twice (Figs. \ref{OH_H2O_slow} and \ref{OH_H2O_high}). When atoms possess kinetic energy exceeding bond energies, bonds dissociate. In high-speed impacts, the atoms in the target receive such large kinetic energy from the impactors that new bonds, such as Si-OH and H$_2$O, are difficult to generate.

Coupling with the interplanetary magnetic fields of solar wind plasma plays an crucial role when the size of the dust particle is smaller than 0.2 $\mu$m \citep{altobelli2019}. While a large mass can contribute to the kinetic energy of a dust particle as its size increases, nano-sized dust particles can also carry substantial energy due to their high velocity. In discussions of space weathering by the solar wind, charged particles such as protons and He$^{+}$ are typically considered. These particles indeed play an important role in space weathering \citep[e.g.,][]{clark2023}. However, nano-dust particles accelerated by the magnetic field of solar wind plasma may also contribute significantly to space weathering on asteroids and the Moon \citep{pieters2016}. Protons, which make up approximately 95\% of solar wind particles, have a mean velocity of $\sim$400 km s$^{-1}$ (its kinetic energy is $\sim$1 keV) at roughly 1-2 AU \citep{gosling2007, nakauchi2021, huang2022}. Compared with charged particles such as H$^{+}$ and He$^{+}$ ions, because a nano-dust particle has a much larger mass, its effect on the removal of OH from phyllosilicates should be larger. In addition, laboratory experiments have shown that solar wind protons can even increase the hydroxyls in phyllosilicate by breaking the Si-O-Si bonds within the tetrahedra of the phyllosilicate structure and binding with oxygen atoms \citep{nakauchi2021}. Accelerated nano-sized dust particles can be an important topic for space weathering, and thus for the surface evolution, of asteroids and the Moon \citep{pieters2016}.

\section{Summary}
Observations of the absorption band at $\sim$2.7 $\mu$m indicate that phyllosilicate minerals on the surface of Ryugu are dehydrated by space weathering, particularly from micrometeoroid bombardment \citep{kitazato2021, matsuoka2023}. To assess the magnitude and process of OH dissociation due to meteoroid impacts, MD simulations using the reactive force field (ReaxFF) potential were conducted. 

When the impactor is not coupled with the interplanetary magnetic fields of solar wind plasma, resulting in average impact velocities of 10-20 km s$^{-1}$, the O-H bonds in serpentine were reduced, especially at $D$ = 2 nm. While the total number of O-H bonds decreased, a portion of the O and H atoms initially connected as Mg-OH was transferred to Si atoms, forming Si-OH. Additionally, free OH and H$_2$O molecules were generated. These newly formed OH can partially offset the overall reduction in O-H bonds. 

When nano-sized dust particles are coupled with the interplanetary magnetic field, the impact velocity increases by an order of magnitude (to $\sim$300 km s$^{-1}$) \citep{juhasz2013, poppe2020}. In this case, even at $D$ = 1 nm, more than 1,000 O-H bonds were dissociated. Nano-dust particles have much greater mass than charged particles such as H$^{+}$ and He$^{+}$ ions. Therefore, in addition to charged particles, nano-dust particles accelerated by the magnetic field of solar wind plasma may also play a role in the space weathering of Ryugu. 

Due to limitations in computational resources, the simulations in this study are constrained in both time and spatial scales. To evaluate the detailed process of space weathering over longer time scales, such as sputtering and evaporation \citep{grayson2022}, a larger volume of the simulation region is necessary, especially for high-speed impact events. We are preparing to expand the simulation environment (size of domain, time steps, and the effect of protons in the solar wind) for future work.

\section*{Acknowledgments}
\noindent
This work was supported by the JAXA Hayabusa2\# International Visibility Enhancement Project. The author appreciates S. C. Chowdhury for offering a ReaxFF potential file. The author also thanks E. Tatsumi, Y. Yokota, K. Honda, M. Ichikawa, S. Murakami, H. Sato, S. Ida, Y. Ochiai for helpful comments. ChatGPT-4o (https://chatgpt.com/) was used for modifications to English grammar.

\bibliography{sample631}{}
\bibliographystyle{aasjournal}




\end{document}